\documentclass[twocolumn]{aastex62}
\usepackage{apjfonts}
\usepackage{here}
\submitjournal{ApJ}

\shorttitle{Flux and Geometry of Charge Exchange in M82}
\shortauthors{Okon et al.}

\begin{document}

\title{Flux Contribution and Geometry of the Charge Exchange Emission in the Starburst Galaxy M82}

\correspondingauthor{Hiromichi Okon}

\author[0000-0002-1671-9904]{Hiromichi Okon}
\affil{Harvard-Smithsonian Center for Astrophysics, 60 Garden Street, Cambridge, MA 02138, USA}

\author[0000-0003-4284-4167]{Randall K. Smith}
\affil{Harvard-Smithsonian Center for Astrophysics, 60 Garden Street, Cambridge, MA 02138, USA}
\author[0000-0003-2251-6297]{Adrien Picquenot}
\affil{Department of Astronomy, University of Maryland, College Park, MD 20742, USA}
\affil{X-ray Astrophysics Laboratory NASA/GSFC, Greenbelt, MD 20771, USA}
\affil{Center for Research and Exploration in Space Science and Technology, NASA/GSFC, Greenbelt, MD 20771, USA}

\author[0000-0003-3462-8886]{Adam R. Foster}
\affil{Harvard-Smithsonian Center for Astrophysics, 60 Garden Street, Cambridge, MA 02138, USA}



\begin{abstract}
Recent X-ray studies of starburst galaxies have found that Charge eXchange (CX) commonly occurs between the outflowing hot plasma and cold gas, possibly from swept-up clouds.
However, the total CX flux and the regions where CX occurs have been poorly understood.
We present an analysis of the {\it XMM-Newton} observations of M82, a prototype starburst galaxy, aiming to investigate these key properties of the CX emisssion.
We have used a blind source separation method in the image analysis with the CCD data which identified a component with the enhanced O-K lines expected from the CX process.
Analyzing the RGS spectra from the region identified by the image analysis, we have detected a high forbidden-to-resonance ratio in the \ion{O}{7} He$\alpha$ triplet as well as several emission lines from K-shell transitions of C, N, and O enhanced in the CX process.
The CX is less responsible for the emission line of Ne and Mg and the accurate estimation of the CX contribution is confirmed to be crucial in measuring chemical abundances.
The temperature of the plasma as electron receiver in the CX process is significantly lower compared to that of the plasma components responsible for most of the X-rays.
From the low temperature and an estimation of the CX emitting volume, we find that the CX primarily occurs in a limited region at the interface of the plasma and gas whose temperature rapidly decreases due to thermal conduction.

\end{abstract}



\section{Introduction} 

Galaxy-scale outflows in starburst galaxies \citep{Veilleux2005,Rubin2014} are responsible for removing energetic and chemically enriched materials from the galaxy, injecting them into the circumgalactic medium (CGM) or even the intergalactic medium \citep{Werk2016,Borthakur2013}.
Such outflows regulate star formation activity within galaxies as well as act as the major driver of cosmic chemical enrichment \citep{Peeples2011,Oppenheimer2006}.
The prevailing picture is that outflows are formed by hot gas shock-heated by supernovae (SNe) and stellar winds that entrain dust and cold gas \citep[e.g.,][]{Chevalier1985}.
X-ray emissions from the multi-phase gas have been widely used as diagnostics tools to investigate their kinematic and chemical properties \citep[e.g.,][]{Tsuru2007,Lopez2020}.

Detection of Charge eXchange (CX) based on recent observations of nearby starburst galaxies with RGS onboard {\it XMM-Newton} \citep{Ranalli2008,Liu2011,Liu2012}, gave us unique insights to the previous studies on outflows.
CX is an atomic process predicted to occur between an ionized plasma and neutral matter in outflows \citep{Lehnert1999,Lallement2004}, and selectively enhances several emission lines such as the forbidden ($f$) and resonance ($r$) in the \ion{O}{7} He$\alpha$ triplet.
The measurements of the flux of the enhanced lines and their ratios allow us to put strong constraints on key parameters (e.g., bulk and turbulence velocities, and density) to understand complicated gas-phase physics (e.g., hydrodynamic effects such as shocks and turbulence, thermal conduction, and non-equilibrium emission processes) that play an important role of the evolution of outflows \citep[e.g.,][]{Strickland2009}.
Although the number of results detecting CX emission in X-ray band is growing, its quantitative flux estimation and the investigation of the CX geometry have been poorly understood, as distinguishing thermal X-ray plasma and CX emission remains difficult and the CX modeling is complex \citep[e.g.,][]{Smith2014}.

M82, the prototype of a starburst galaxy, is where CX emission was detected for the first time \citep{Ranalli2008}.
Because of its proximity \citep[3.6~Mpc][]{Freedman1994} and large inclination angle \citep[$i\sim80^\circ$;][]{McKeith1995}, M82 is regarded as the ideal target for observational studies on outflows.
Biconical outflows from the nuclear region of M82 extend perpendicularly to the galactic disc and their properties have been widely studied across the electromagnetic spectrum, tracing atomic \ion{H}{1} and molecular gas \cite[e.g.,][]{Walter2002,Salak2013}, warm-ionized gas in H$\alpha$ \cite[e.g.,][]{McKeith1995,Ohyama2002}, hot plasma emitting X-rays \citep[e.g.,][]{Watson1984,Lopez2020}, and dust in the UV, IR, and submillimeter bands \cite[e.g.,][]{Origlia2004,Hoopes2005}.
Past observations of M82 aiming to investigate the nature of X-rays pointed out the discrepancy between the plasma model and the CCD spectra in the \ion{O}{7} He$\alpha$ triplet in both the galactic disk and outflow regions of M82 \citep{Konami2011,Lopez2020}.

Here we report on an analysis of {\it XMM-Newton} data of M82, combining the application of a blind source separation method together with high resolution spectroscopy to pin down the origin of the CX emission.
Throughout this paper, we adapt 3.6~Mpc as the distance to M82 \citep{Freedman1994}, and the statistical errors are quoted at the 1$\sigma$ level.

\section{Observations and Data Reduction}

\begin{deluxetable*}{cccccccc}
\tablecaption{Observation log \label{table1}}
\tablecolumns{6}
\tablewidth{0pt}
\tablehead{
\colhead{Target} &
\colhead{Obs. ID} &
\colhead{Obs. Date} &
\colhead{(R.A., Dec.)\tablenotemark{a}} &
\colhead{Roll angle~(deg)} &
\multicolumn3c{Effective Exposure (ks)} \\
&&&& & MOS\tablenotemark{b} & pn & RGS
}
\startdata
	M82 & 0112290201 & 2001 May 6 & ($9^{\rm h} 55^{\rm m} 49\fs99,~+69\degr 40\arcmin 45\farcs0$) & 293.0 & 46.7 &18.0 & 24.7 \\ 
	M82 & 0206080101 & 2004 April 23 & ($9^{\rm h} 55^{\rm m} 52\fs20,~+69\degr 40\arcmin 47\farcs0$) & 319.3 & 115.0 & 46.1 & 71.5 \\
	M82~ULX & 0560590101 & 2008 October 3 & ($9^{\rm h} 55^{\rm m} 50\fs19,~+69\degr 40\arcmin 47\farcs0$) & 138.3 & 52.3 & 28.1 & 30.9 \\
	M82~ULX & 0560590201 & 2009 April 17 & ($9^{\rm h} 55^{\rm m} 50\fs19,~+69\degr 40\arcmin 47\farcs0$)  & 315.7 & 29.1 & 15.1 & 21.9 \\
	M82~ULX & 0560590301 & 2009 April 29 & ($9^{\rm h} 55^{\rm m} 50\fs19,~+69\degr 40\arcmin 47\farcs0$) & 295.9 & 33.0 & 13.4 & 30.1 \\
	M82~X-1 & 0657802101 & 2011 September 24 & ($9^{\rm h} 55^{\rm m} 49\fs91,~+69\degr 40\arcmin 44\farcs4$) & 147.2 & 25.3 & 6.5 & 17.4 \\
	M82~X-1 & 0657802301 & 2011 November 21 & ($9^{\rm h} 55^{\rm m} 49\fs91,~+69\degr 40\arcmin 44\farcs4$)  & 100.3 & 20.8 & 9.2 & 18.7  \\
	M82~X-2 & 0870940101 & 2021 April 16 & ($9^{\rm h} 55^{\rm m} 51\fs03,~+69\degr 40\arcmin 45\farcs0$)  & 312.6 & 41.8 & -- & 27.2 \\
	M82~X-2 & 0870940401 & 2021 April 16 & ($9^{\rm h} 55^{\rm m} 51\fs03,~+69\degr 40\arcmin 45\farcs0$)  & 304.8 & 60.0 & 25.1 & 31.9 \\
\enddata
\label{tab:use_data}
\tablenotetext{a}{Equinox in J2000.0.}
\tablenotetext{b}{The total effective exposure of MOS~1 and MOS~2.}
\end{deluxetable*}

\begin{figure*}[ht]
\begin{center}
\includegraphics[width=15cm]{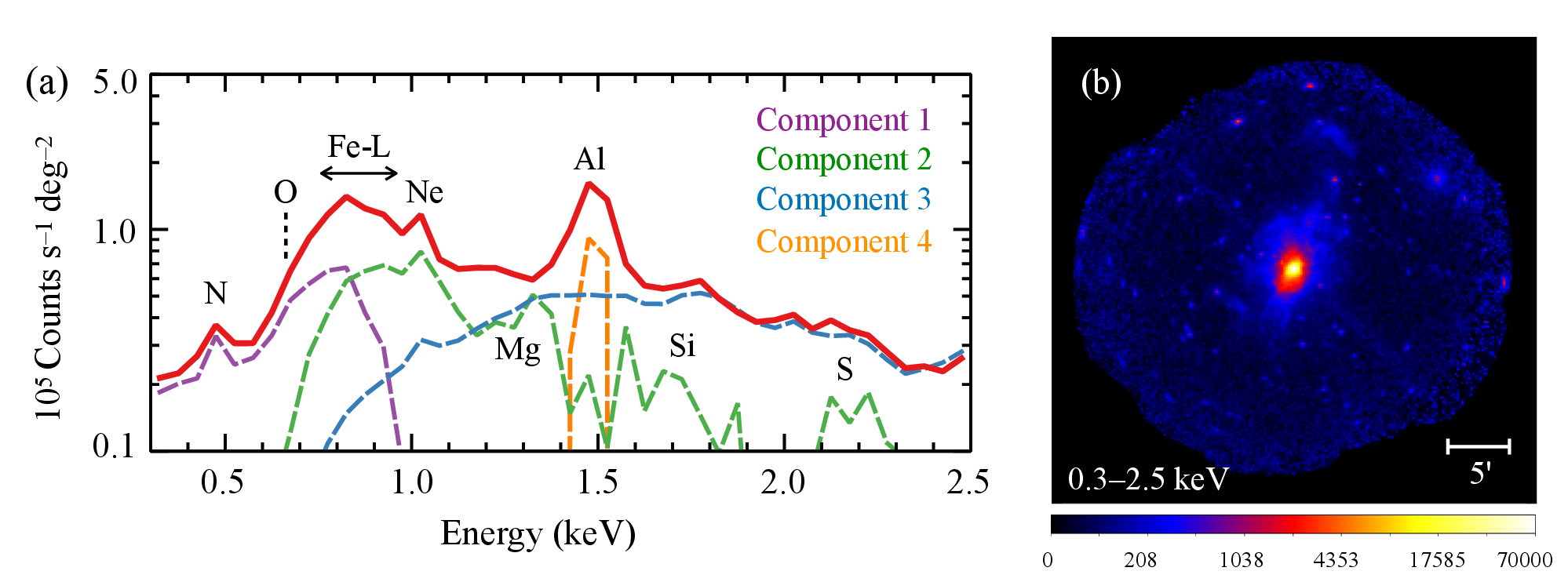} 
\\
\vspace{4mm}
\includegraphics[width=15cm]{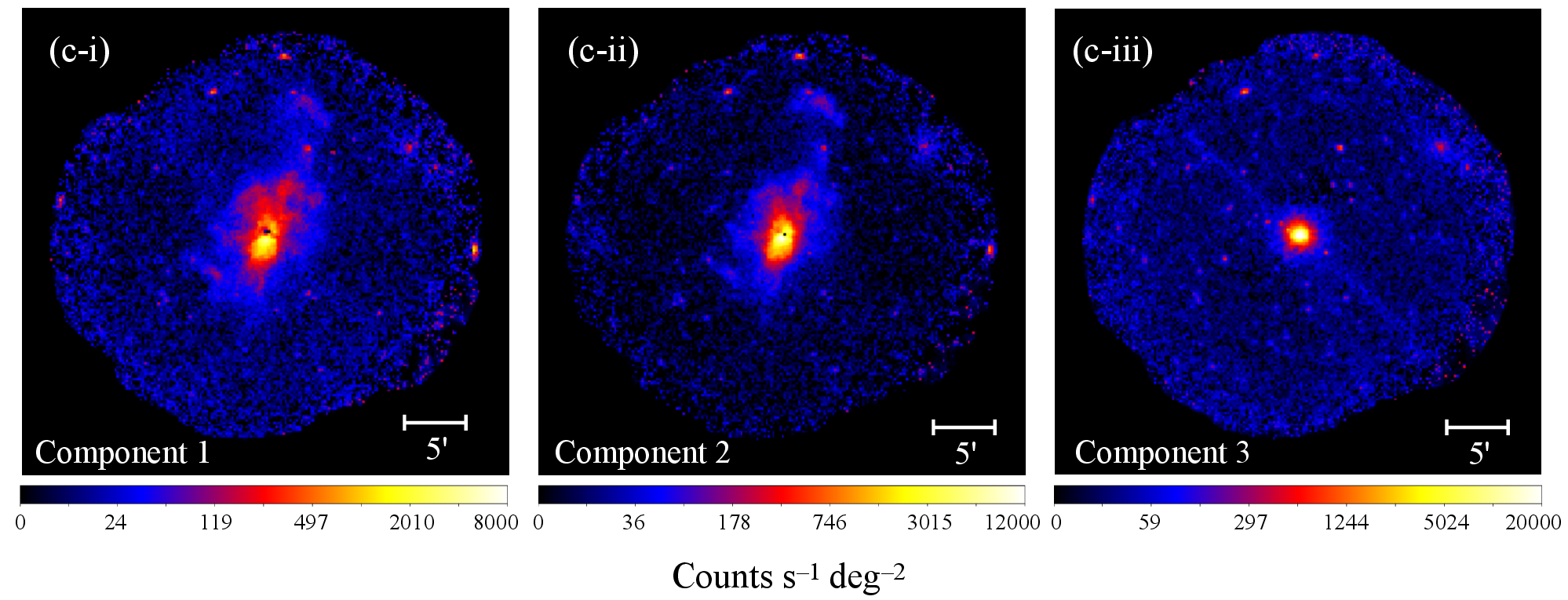} 
\\
\vspace{3mm}
\includegraphics[width=15cm]{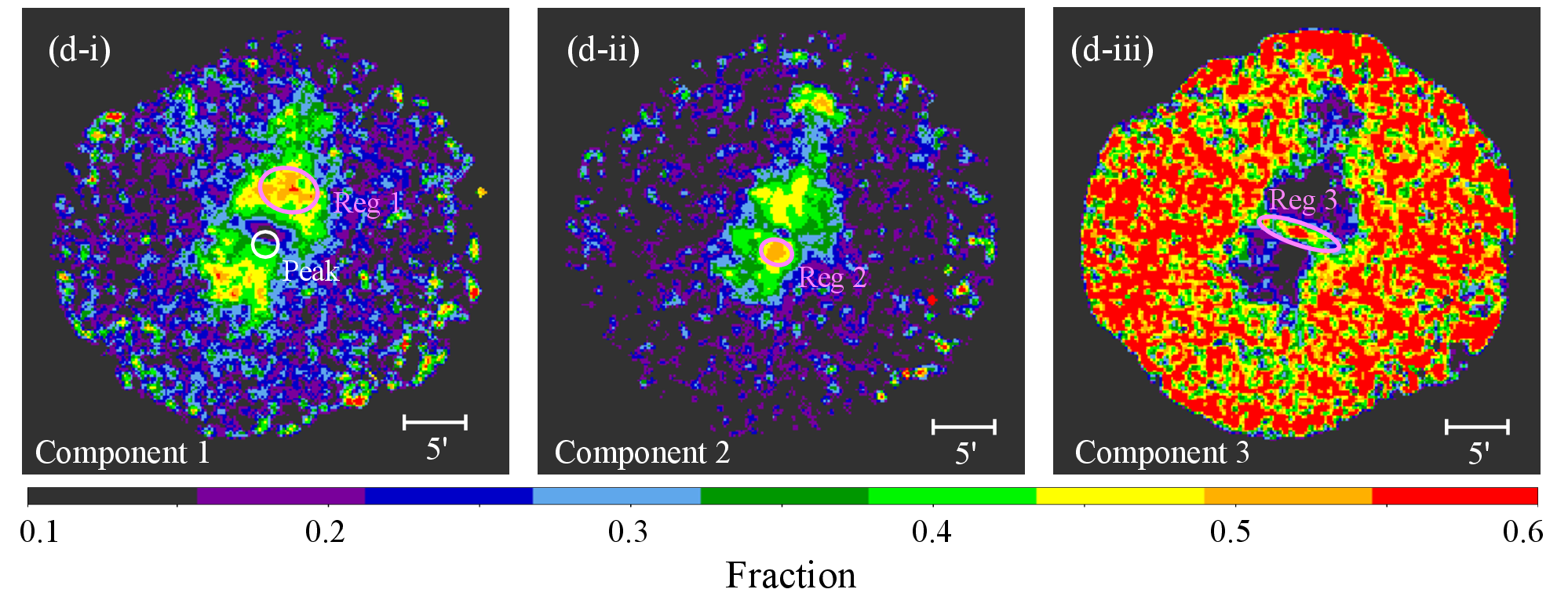} 
\end{center}
\vspace{-2mm}
\caption{
Results of the GMCA analysis applied to the M82.
(a) Output spectra of the four components.
Components~1~(purple), 2~(green), 3~(blue), and 4~(orange) predominantly consist of CX, hot plasma, nontherml emission, and line emission due to detector background, respectively.
The input data before the component extraction~(red) is given by the sum of the components.
Counts maps of (b) the input and (c-i)-(c-iii) the components.
(d-i)-(d-iii) Fraction of each component, defined as $f_i$ = $N_i / \sum_{i=1-4} N_i$, where $N_i$ is the number of photons in the corresponding count map.
The regions enclosed by the magenta lines in panel (d) are the three representative regions whose spectra are plotted in Figure~\ref{fig2}.
The white circle indicates the Peak region as shown in Figures~\ref{fig3}(a).
}
\label{fig1}
\end{figure*}

\begin{figure}[ht]
\begin{center}
\vspace{-0mm}
\includegraphics[width=7.5cm]{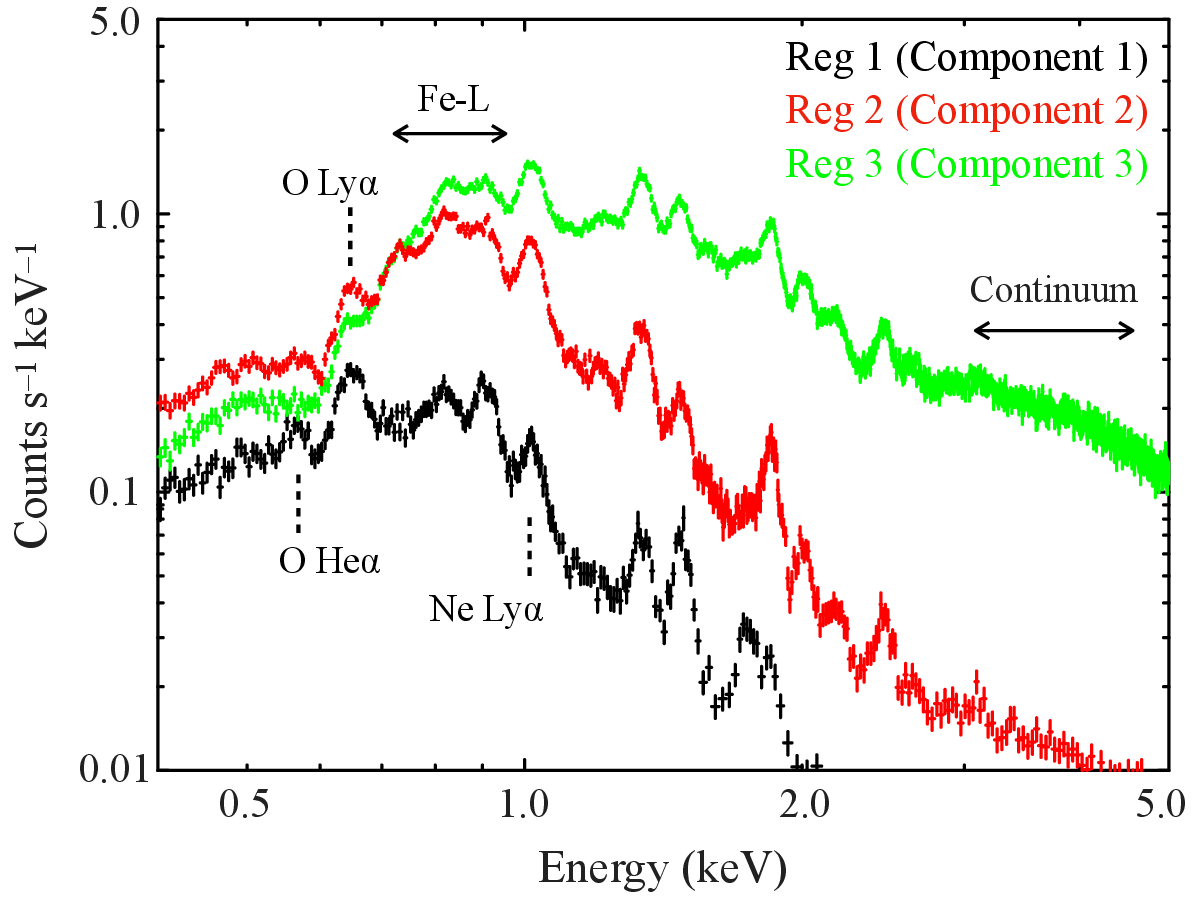} 
\end{center}
\vspace{-5mm}
\caption{
MOS1$+$2 spectra extracted from the representative region 1 (black), 2 (red), and 3 (green) in Figure~\ref{fig1}(c).
}
\label{fig2}
\end{figure}

\begin{figure*}[ht]
\begin{center}
\vspace{-0mm}
\includegraphics[width=17cm]{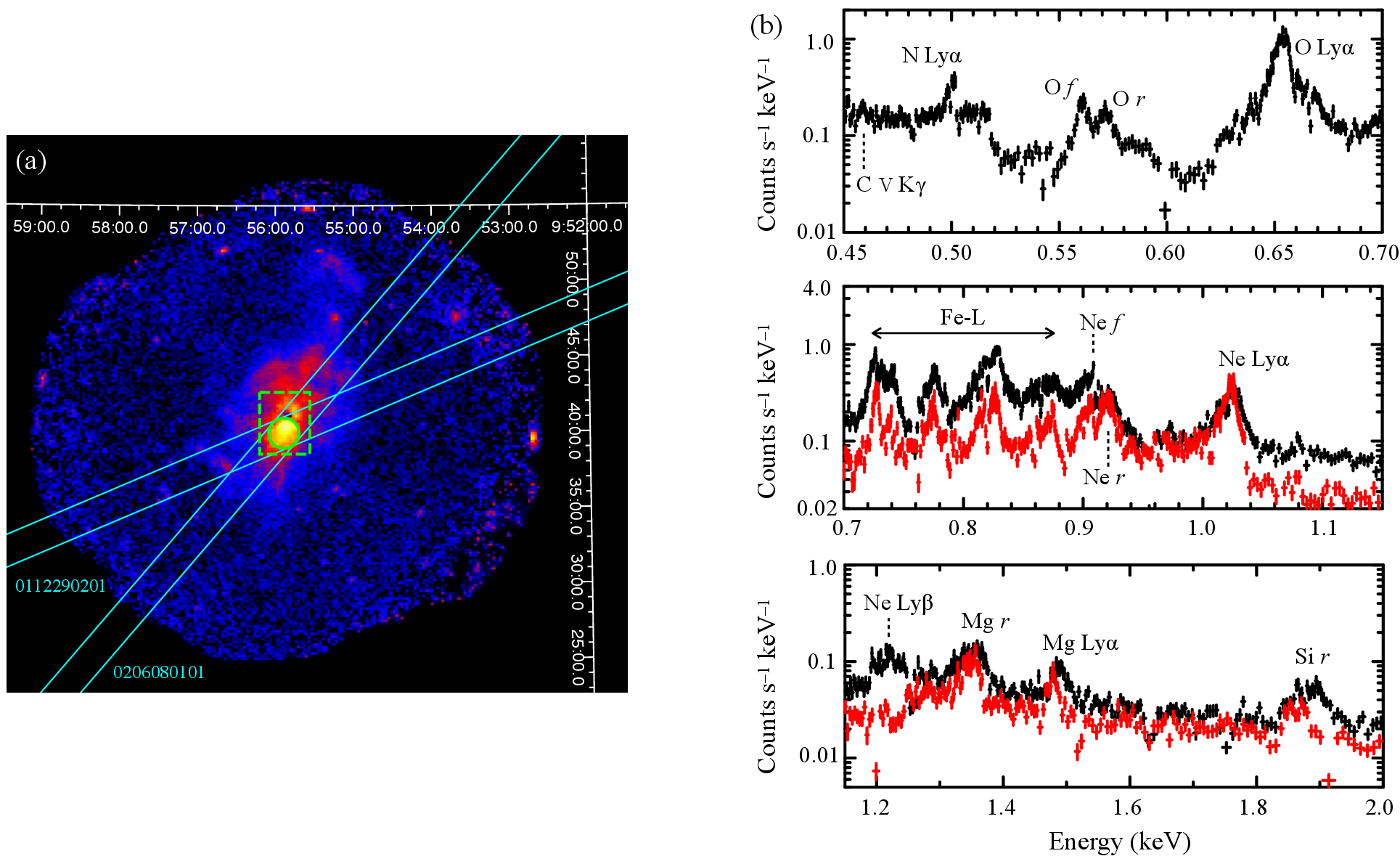} 
\end{center}
\vspace{-5mm}
\caption{
(a) Same as Figure~\ref{fig1}(b-i) but we show the location of Peak region (green circle) and RGS spectral extraction regions (cyan) of OBS. ID\,=\,0112290201 and 0206080101 whose the roll angles are the maximum and minimum, respectively.
The green box displays the area of Figure~\ref{fig7}.
(b) RGS1$+$2 first- (black) and second-order (red) spectra where all used data sets are integrated.
}
\label{fig3}
\end{figure*}

\begin{figure}[ht]
\begin{center}
\vspace{-0mm}
\includegraphics[width=7.5cm]{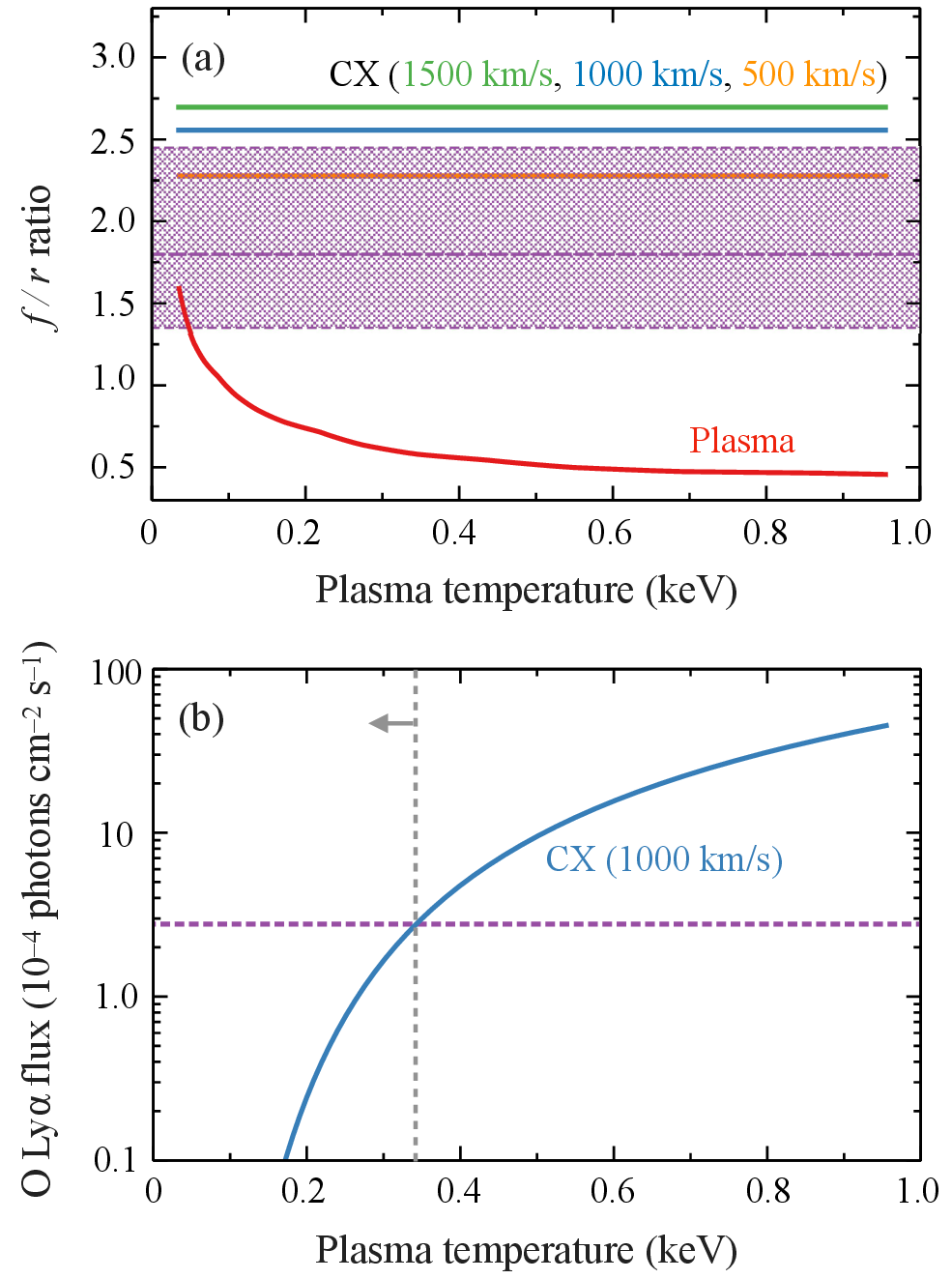} 
\end{center}
\vspace{-5mm}
\caption{
Comparison between the observational and theoretical (a) \ion{O}{7} $f/r$ ratio and (b) \ion{O}{8} Ly$\alpha$ flux.
The purple area (line) indicates the observed line ratio (flux). 
The red curves in panel (a) is the ratio predicted from collisional ionization equilibrium plasma as a function of the plasma temperature.
The orange, green, and blue curves shows the ratios in the CX cases of $\upsilon_{\rm col}=500~{\rm km/s}$, $1000~{\rm km/s}$, and $1500~{\rm km/s}$, respectively.
In panel (b), because three curves completely overlap each other, we only plot the case of $\upsilon_{\rm col}=1000~{\rm km/s}$.
The gray dot line and arrow represent the constrained upper limit of $kT_{e,\,{\rm CX}}=0.35~{\rm keV}$.
}
\label{fig4}
\end{figure}

\begin{figure*}[ht]
\begin{center}
\vspace{-0mm}
\includegraphics[width=17cm]{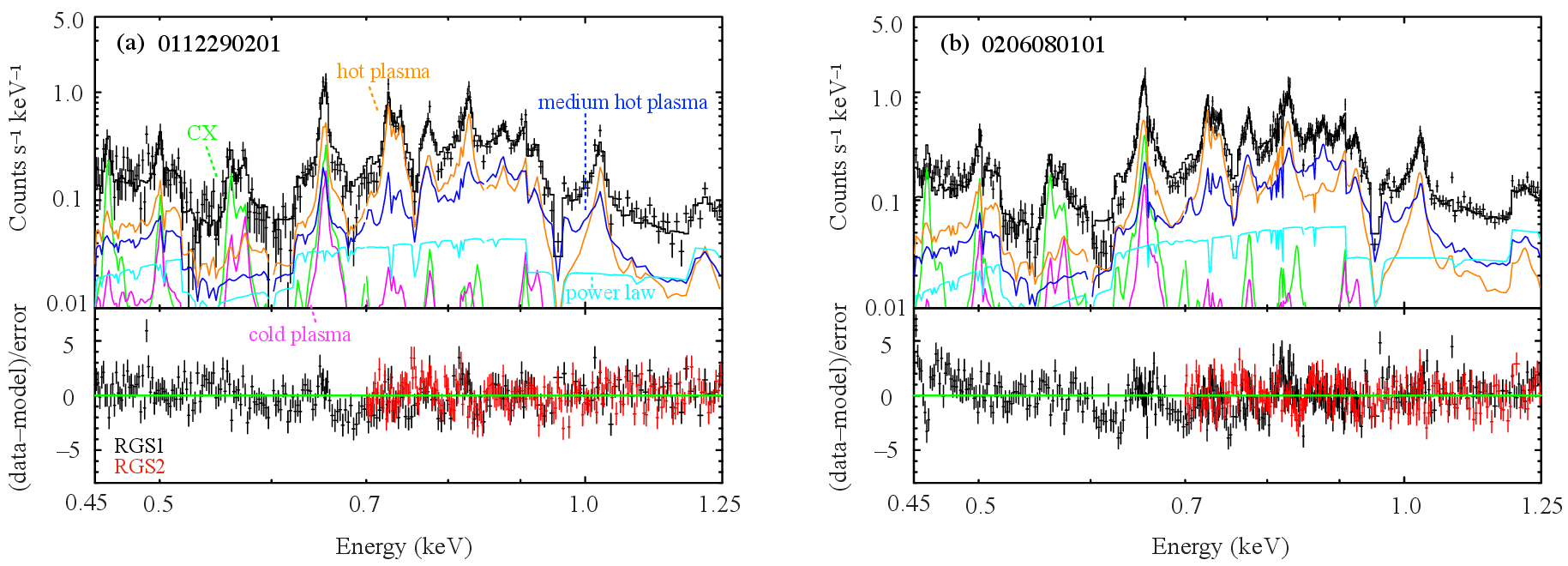} 
\includegraphics[width=17cm]{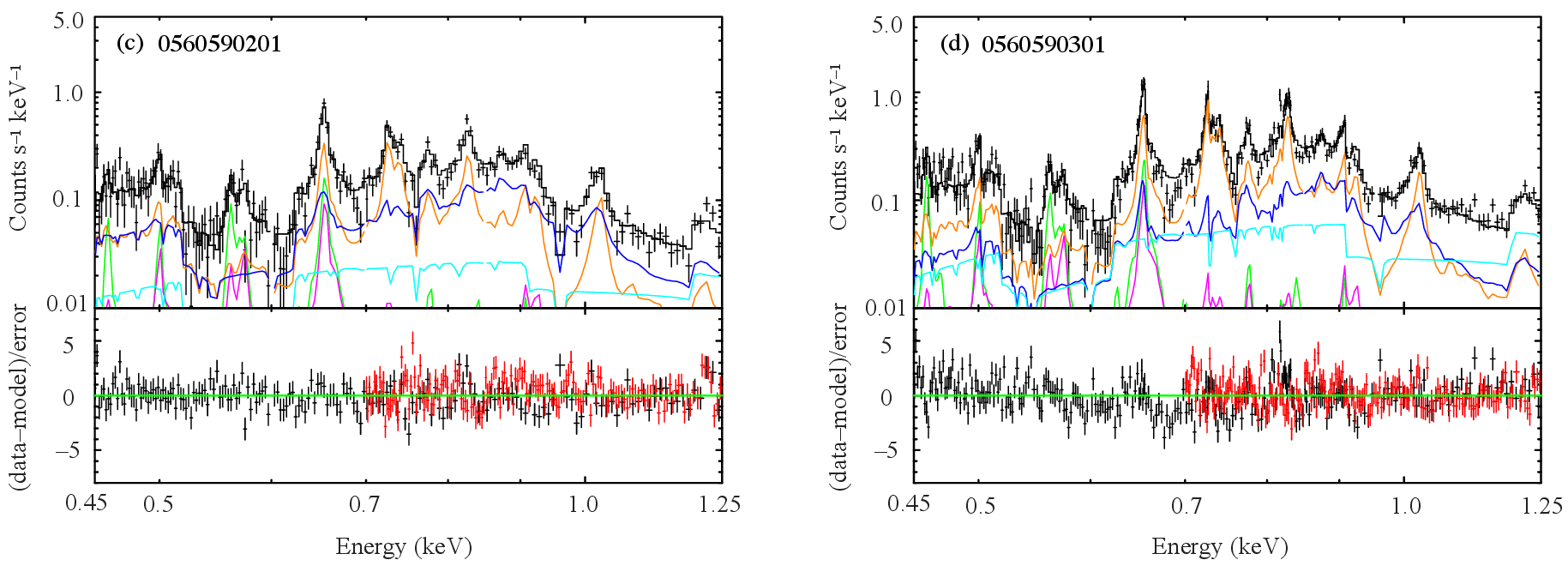} 
\includegraphics[width=17cm]{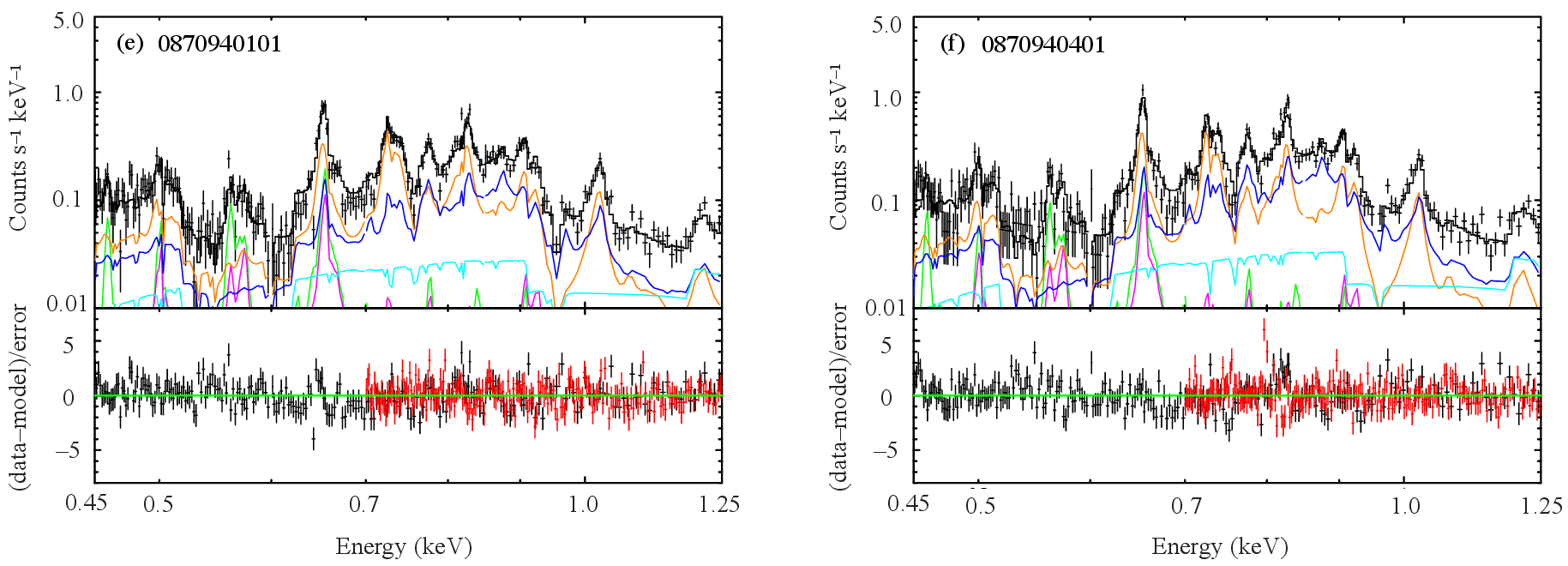} 
\end{center}
\vspace{-5mm}
\caption{
(a) RGS1$+$2 first- (black) and second-order (red) spectra of OBS.ID$=$0112290201 plotted with the best-fit model.
The blue, magenta, orange, green, and black curves represent the CX model, the cold, medium hot, hot plasma models, and the sum of the models to RGS1$+$2 first-order data.
(b)--(f) Same as panel (a) but for OBS. ID$=$0206080101, 0206080101, 0560590201, 0560590301, 0870940101, and 0870940101.
}
\label{fig5}
\end{figure*}

M82 has been observed with {\it XMM-Newton} many times between 2003 to 2021.
In most of these observations, the field of view (FoV) of MOS and pn CCDs completely covers the entirety of M82 including both outflows.
We used MOS and pn data only for the image analysis and identifying spectral characteristics.
We reprocessed this data with the Science Analysis System software (SAS) version 19.1.0 and the Current Calibration Files, following the cookbook for analysis procedures of extended sources\footnote{https://heasarc.gsfc.nasa.gov/docs/xmm/esas/cookbook/xmm-esas.html}.
We discarded short time Observation IDs whose MOS1+2 exposure time after the screening is less than 20~ks.
The observations that meet the criteria are summarized in Table~\ref{table1}.
Furthermore, two datasets (Obs. ID = 0870940101 and 0870940401) observed in the small window mode were not used.
We used RGS data only for spectral analysis.
These data were reprocessed using the {\tt rgsproc} task in SAS.
We extracted background light curves and filter out observation periods when the background count rate is higher than 0.15~cnt\,s$^{-1}$. 

\section{Analysis and Results} \label{sec:floats}

\subsection{Imaging Analysis} \label{subsec:ia}

Our first purpose is to constrain the spatial distribution of a component with the enhanced \ion{O}{7} He$\alpha$ triplet expected from the CX process \cite[e.g.,][]{Lallement2004}.
We used a blind source separation method, the General Morphological Components Analysis (GMCA) \citep{Bobin2015}, that was designed specifically for cosmic microwave background reconstruction using {\it Planck} data.
The method was recently introduced to an analysis of {\it Chandra} data by \cite{Picquenot2019} and the authors demonstrated that this algorithm works as a powerful tool to isolate physically meaningful components in the data, and identify extraction regions of interest.
In the case of M82, the observed X-rays arise from multi-temperature thermal plasmas (cold:\,$\sim0.2$~keV, medium hot:\,$\sim0.5$~keV, hot:\,$\sim0.9$~keV, and extreme hot:\,$\leq6$~keV) as well as nonthermal emission, in addition to the CX emission we are interested in \citep[e.g.,][]{Konami2011,Lopez2020}.
Although the GMCA has no intrinsic knowledge of these different components, if the CX emission has an identifiable spectral and spatial shape in the data, it should be able to extract it.

The primary concept of the GMCA method is to take into account the morphological particularities of distinct components by measuring the sparsity level in the wavelet domain for each energy slice of a 3-D data cube of the photon position ($x$,~$y$) and energy ($E$).
Inputs to the algorithm are the data cube and the user-defined number $N$ of components to extract, and the output is a set of $N$ images associated with the spectra.
In our analysis, the position ($x$,~$y$) and energy ($E$) information in the cube correspond to the sky coordinate of each event on band images and the energy range we define, respectively.
We used the data in the energy range from 0.3~keV to 2.5~keV where all CX-enhanced lines (e.g., \ion{C}{6}~He$\gamma$:\,0.459~keV; \ion{N}{7}~Ly$\alpha$:\,0.500~keV... \ion{Si}{13}~He$\alpha$~$f$:\,1.839~keV) discussed in previous work \citep[e.g.,][]{Zhang2014} are covered, and generated 44 band images with ${\tt binning=3}$ in equal 0.05 keV increments, with {\tt adapt-merge}.
Subsequently, we merged them together to build the input data cube.
The number $N$ was fixed to 4.
For the case of $N\geq5$, two of the components show similar images and spectra, which can be interpreted as the overfitting of the data.
On the other hand, for $N\leq3$, the CX component of interest and other components cannot be disentangled well.

Figure~\ref{fig1}(a) shows the spectra of the primary extracted components.
The spectrum of Component~1 is characterized by the soft emission below 1~keV including the \ion{O}{7} He$\alpha$ triplet selectively enhanced by the CX process as reported by previous work \citep[e.g.,][]{Ranalli2008}.
In the spectrum of Component~2, one can see a strong Fe-L complex and \ion{Ne}{8} Ly$\alpha$ lines.
These characteristics are consistent with the interpretation that Component 2 mainly represents thermal emission from moderately hot plasma ($kT_e\sim$~0.5--1.0~keV).
The spectrum of Component~3 is described by a featureless spectral continuum dominant above $\sim2$~keV.
Given the count map as shown in Figure~\ref{fig1}(b), Component~3 can be considered to be dominated by a nonthermal emission from M82 X-1 \citep{Konami2011}.
Component~4 consists of a line-like structure at $\sim$1.5~keV.
The structure arises from a neutral Al line (1.49~keV) in the instrumental background \citep{Kuntz2008} whose careful handling is often required in the analyses of diffuse sources \cite[e.g.,][]{Okon2020,Okon2021}.

\subsection{CCD Spectral Characteristics} 

The spectral interpretations in \S\ref{subsec:ia} are, however, merely qualitative.
To quantify these interpretations we used MOS1$+$2 spectra extracted from representative regions in Figure~\ref{fig1}(d).
Regions~1--3 are dominated by GMCA components 1--3 based on the fraction map in Figure~\ref{fig1}(c), defined as $f_i$ = $N_i / \sum_{i=1-4} N_i$, where $N_i$ is the number of photons in the count map for the component $i$ in Figure~\ref{fig1}(b).
We found that the MOS spectrum from Region~1 exhibit the strong \ion{O}{7} He$\alpha$ and \ion{O}{8} Ly$\alpha$ lines expected from the GMCA results (Figure~\ref{fig2}).
We also confirmed that the enhanced Fe-L complex and \ion{Ne}{9}~Ly$\alpha$ from a hot thermal plasma is seen in spectra from Regions~2 and that a nonthermal continuum above 2~keV dominates the spectrum from Region~3.

\subsection{RGS Spectral Analysis \label{sec:SA}}

\begin{deluxetable*}{cccccccc}[ht]
\tablecaption{The intensities of the O and Ne emission. \label{table2}}
\tablehead{
\colhead{Obs. ID} &
\colhead{{O~$f$}}$^a$ & 
\colhead{O~$r$}$^a$ &
\colhead{O~Ly$\alpha$}$^a$ &
\colhead{O~$f$\,/\,O $r$} &
\colhead{Ne~$f$}
} 
\startdata
       0112290201 & 1.08$^{+0.18}_{-0.17}$ & 0.62$^{+0.17}_{-0.16}$ & 3.12$\pm$0.15 & 1.73$^{+0.64}_{-0.43}$ & $0.63\pm0.10$ \\ 
       0206080101 & 0.91$\pm$0.09 & 0.51$^{+0.09}_{-0.08}$ & 3.05$\pm$0.10 & 1.78$^{+0.57}_{-0.41}$ & $0.93\pm0.08$ \\ 
       0560590201 & 0.84$^{+0.14}_{-0.13}$ & 0.47$^{+0.12}_{-0.19}$ & 2.52$^{+0.19}_{-0.13}$ & 1.79$^{+0.63}_{-0.59}$ & $0.66\pm0.12$ \\ 
       0560590301 & 0.98$\pm$0.20 & 0.54$^{+0.21}_{-0.18}$ &  2.88$\pm$0.21 & 1.81$^{+0.71}_{-0.42}$ & $0.71\pm0.10$ \\ 
       0870940101 & 0.83$^{+0.16}_{-0.15}$ & 0.46$^{+0.15}_{-0.14}$ & 2.17$\pm$0.13 & 1.80$^{+0.78}_{-0.51}$ & $0.70\pm0.10$\\ 
       0870940401 & 0.77$^{+0.16}_{-0.15}$ & 0.40$^{+0.15}_{-0.14}$ & 2.45$^{+0.14}_{-0.13}$ & 1.93$^{+0.65}_{-0.44}$ & $0.79\pm0.12$ \\ \hline
       Exposure Weighted & 0.90$\pm$0.14 & 0.50$\pm$0.14 & 2.77$^{+0.15}_{-0.14}$ & 1.80$^{+0.65}_{-0.45}$ & $0.79\pm0.10$ \\ 
\enddata
\tablenotetext{a}{In units of $10^{-4}$ photons cm$^{-2}$ s$^{-1}$.}
\end{deluxetable*}


\begin{deluxetable*}{llcccccc}[ht]
\tablecaption{Best-fit model parameters of the spectra of the all observatiion.\label{table3}}
\tablehead{
\colhead{Model Function} &
\colhead{Parameters} &
\multicolumn6c{Obs. ID}  \\
& &
\colhead{0112290201} & 
\colhead{0206080101} &
\colhead{0560590201} &
\colhead{0560590301} &
\colhead{0870940101} &
\colhead{0870940401}
} 
\startdata
       TBabs & $N_{\rm H,\,MW}$ (10$^{21}\,{\rm cm^{-2}}$) & \multicolumn6c{$=0.4$ (fixed)} \\ 
       TBvarabs & $N_{\rm H,\,M82}$ (10$^{21}\,{\rm cm^{-2}}$) & $0.8\pm0.2$ & 1.78$\pm$0.07 & $0.6\pm0.2$ &$1.0\pm0.1$ & $0.9\pm0.1$ & $1.2\pm0.1$ \\ 
       VACX2 & $kT_{e,{\rm CX}}$ (keV) & \multicolumn6c{$=kT_{e,\,{\rm cold}}$} \\
       & $Z_{\rm C}$ (Solar) & $2.6\pm{0.5}$ & 1.8$\pm$0.3 & 0.8$\pm$0.3 & 2.7$\pm$0.5 & $1.1\pm0.3$ & $1.6\pm0.5$ \\
       & $Z_{\rm N}$ (Solar) & $0.8\pm{0.2}$ & 0.79$^{+0.07}_{-0.06}$ &0.6$\pm$0.2  & $0.9\pm{0.1}$ & $0.8\pm0.1$ & $0.8\pm0.1$ \\
       & $Z_{\rm O}$ (Solar) & $0.42\pm0.06$ & 0.38$\pm$0.01 & 0.34$\pm$0.03 & $0.43\pm0.03$ & $0.37\pm0.02$ & $0.42\pm0.03$ \\
       & $Z_{\rm Ne}$ (Solar) & $1.1\pm0.3$ & 0.99$\pm$0.06& $1.1\pm0.2$ & $1.2\pm0.1$ & $1.1\pm0.1$ & $1.2\pm0.1$ \\
       & $Z_{\rm Fe}$ (Solar) & $0.40^{+0.05}_{-0.06}$ &0.37$\pm$0.03 & 0.26$\pm$0.04 & $0.35^{+0.04}_{-0.03}$ & $0.28\pm0.03$ & $0.31\pm0.03$ \\
       & $Z_{\rm Ni}$ (Solar) & 0.9$\pm0.3$ &0.9$\pm$0.1 &0.7$\pm$0.3 & $0.9\pm0.2$ & $0.6\pm0.3$ & $0.8\pm0.3$\\
       & $\upsilon_{\rm col}~{\rm (km/s)}$ & \multicolumn6c{1000~(fixed)} \\
       & ${\rm VEM_{CX}}^{a}$ (10$^{57}\,{\rm cm^{-5}}$) & $0.5\pm0.2$ & $1.5\pm0.3$ & $0.6\pm0.2$ & $0.7\pm0.2$ & $0.6\pm0.2$ &  $0.7\pm0.2$ \\  
       bvapec & $kT_{e,\,{\rm cold}}$ (keV) & $0.22\pm0.02$ & $0.21\pm0.01$ & $0.22\pm0.02$ & $0.20\pm0.02$ & $0.20\pm0.02$ & $0.22\pm0.02$ \\
       & Abundance & \multicolumn6c{$=$ Abundances of the ACX component} \\
       & $\sigma_{\rm cold}~{\rm (km/s)}$ & \multicolumn6c{0 (fixed)}  \\
       & $\upsilon_{\rm cold}~{\rm (km/s)}$ & \multicolumn6c{0 (fixed)}  \\
       & ${\rm VEM_{cold}}^{b}$ (10$^{61}\,{\rm cm^{-5}}$) & $\leq0.7$ & $\leq0.9$ & $\leq0.6$ & $\leq06$ & $\leq0.5$ & $\leq0.5$\\
       bvapec & $kT_{e,\,{\rm med}}$ (keV) & $0.44\pm0.01$ & $0.419\pm0.006$ & $0.44\pm0.02$ & $0.45\pm0.01$ & $0.44\pm0.01$ & $0.41\pm0.01$\\
       & Abundance & \multicolumn6c{$=$ Abundances of the ACX component} \\
       & $\sigma_{\rm med}~{\rm (km/s)}$  & $\leq558$ & $796^{+173}_{-184}$ & $581^{+157}_{-149}$ & $345^{+63}_{-61}$ & $1183^{+163}_{-212}$ & $858^{+179}_{-176}$ \\
       & $\upsilon_{\rm med}~{\rm (km/s)}$ & $313^{+46}_{-44}$ & $320^{+60}_{-66}$ & $260^{+118}_{-106}$ & $261^{+128}_{-129}$ & $1046^{+138}_{-168}$ &  $970^{+168}_{-170}$ \\
       & ${\rm VEM_{med}}^{b}$ (10$^{61}\,{\rm cm^{-5}}$) & $2.9\pm0.5$ &  $5.4\pm0.4$  & 2.9$\pm$0.5 & $3.8\pm0.4$ &  $3.3^{0.6}_{-0.5}$ & $3.2\pm0.5$ \\
       bvapec & $kT_{e,\,{\rm hot}}$ (keV) & $0.85\pm0.03$ & $0.82\pm0.01$ & 0.89$\pm$0.03 & $0.91\pm0.02$ & $0.81\pm0.03$ & $0.79\pm0.03$\\
       & Abundance & \multicolumn6c{$=$ Abundances of the ACX component} \\
       & $\sigma_{\rm hot}~{\rm (km/s)}$ & $\leq2149$ & $122^{+23}_{-41}$ & $2061^{+527}_{-630}$ & $799^{+258}_{-267}$ & $\leq1429$ & $\leq891$\\
       & $\upsilon_{\rm hot}~{\rm (km/s)}$ & $670^{+273}_{-277}$ & $-114^{+92}_{-95}$ & $1004^{+340}_{-312}$ & $\leq 900$ & $189^{+172}_{-169}$ & $99^{+153}_{-161}$ \\
       & ${\rm VEM_{hot}}^{b}$ (10$^{61}\,{\rm cm^{-5}}$) & $2.2\pm0.3$ & $4.6\pm0.3$ & 3.5$\pm$0.5 & $2.6^{+0.2}_{-0.3}$ & $2.6^{+0.4}_{-0.3}$ & $3.1\pm0.4$ \\
       power law & $\Gamma$ & \multicolumn6c{0.55 (fixed)} \\
       & Norm$^{c}$ & $0.8\pm0.1$ & $1.33\pm0.07$ & 0.7$\pm$0.1 & $1.2\pm0.1$ & $0.6\pm0.1$ & $0.7\pm0.1$\\ \hline
      & $C$-stat\,/\,d.o.f. & 3986\,/\,3261 & 4153\,/\,3252 &2990\,/\,3252 & 4093\,/\,3252 & 3867\,/\,3252 & 3846\,/\,3252 \\
\enddata
\tablenotetext{a}{The volume emission measure integrated over the line of sight, i.e., $\int n_e\,n_{\rm (H+He)}\,dV$.}
\tablenotetext{b}{The volume emission measure integrated over the line of sight, i.e., $\int n_e\,n_{\rm H}\,dV$.}
\tablenotetext{c}{The unit is $10^{-4}$ photons\,/\,keV\,/\,cm$^2$/s at 1 keV}
\end{deluxetable*}

Unfortunately, CCDs cannot spectrally-resolve the \ion{O}{7} triplet to detect conclusive evidence of CX via a significantly high $f/r$ ratio. 
We therefore turned to RGS spectra.
Given the deviation of the wavelength of the incident photon against its off-axis angle\footnote{https://heasarc.gsfc.nasa.gov/docs/xmm/esas/cookbook/xmm-esas.html}, the difference of the center energy of the \ion{O}{7} $f$ and $r$ corresponds to the source size of $\sim4'$.
We focused on a bright peak with an angular diameter $\sim2'$ enough to resolve the multiplet lines, in the south outflow as shown in the map of Component~1 (noted as ``Peak'' in Figure~\ref{fig3}(a)).
We used all data sets in which the RGS FoVs completely cover the Peak region regardless of the roll angle (Obs. ID$=$0112290201, 0206080101, 0560590201, 0560590301, 0870940101, 0870940401).
Figure~\ref{fig3}(b) shows the first- and second-order spectra of the Peak region, where the data from all the observations are integrated to improve the photon statistics.
The $f$ and $r$ line in the RGS data are clearly resolved and the high $f/r$ ratio is confirmed.

We first measured the line intensity of the \ion{O}{7}~He$\alpha$~$f$ and $r$ lines, as well as \ion{O}{8}~Ly$\alpha$
since their ratios offer helpful guides in complex CX modeling.
Given the variation in the roll angle among the datasets, we independently fitted RGS1$+$2 first-order spectra in the energy band of 0.55--0.70~keV with a phenomenological model consisting of a bremsstrahlung continuum and five Gaussians accounting for \ion{O}{7}~$f$ (0.574~keV), intermediate (0.569~keV), and $r$ (0.561~keV), \ion{O}{8}~Ly$\alpha$ (0.654~keV), and \ion{O}{7}~Ly$\beta$ lines (0.666~keV).
The central energy of the Gaussian components was fixed to the corresponding line energy.
The intensity of the Gaussians and the parameters concerning the continuum were allowed to vary.
We convolved RGS response matrices (RMFs) with the spatial profile of source emission in the MOS$+$pn image in the 0.55--0.70~keV band, using the {\tt FTOOL} {\tt ftrgsrmfsmooth}. 
In the spectral fitting, we used XSPEC software version 12.11.1 \citep{Arnaud1996} and the $C$-statistic \citep{Cash1979} on unbinned spectra.
The results are summarized in Table~\ref{table2}.
Finally, we calculated the exposure-weighted mean intensity~$I$ and the statistical error $\sigma_{I,\,{\rm stat}}$ among the measurements as follows, 
\begin{equation}
I = \frac{\sum_i I_i \times {t_i}}{\sum_i t_i},  \label{eq1}
\end{equation}
\begin{equation}
\sigma_{I,\,{\rm stat}} = \sqrt{\frac{\sum_i {\sigma_{I_i,\,{\rm stat}}}^2 \times t_i }{\sum_i  t_i}} \label{eq2}
\end{equation}
where $I_i$, $\sigma_{I_i,\,{\rm stat}}$, and $t_i$ are the intensity, statistical error, and exposure time of the observation~$i$, respectively.

Figure~\ref{fig4}(a) compares the measured \ion{O}{7}~He$\alpha$ $f/r$ ratio with theoretically expected curves from plasma and CX models computed with {\tt PyatomDB}\footnote{https://atomdb.readthedocs.io/en/master/} and the AtomDB CX package\footnote{https://github.com/AtomDB/ACX2}  based on cross sections from the Kronos database \citep{Mullen2016,Mullen2017,Cumbee2018}, respectively.
Here, we presented three cases of the CX process where the collision velocity between the plasma and gas is $500~{\rm km/s}$, $1000~{\rm km/s}$ or $1500~{\rm km/s}$, taking into account the velocity of outflow plasma estimated with the \cite{Chevalier1985} model $\sim$1000--2500~km/s \citep{Strickland2009} and the observed H$\alpha$ clump $\sim$600~km/s \citep{Shopbell1998}.
We found that the observed $f/r$ ratio requires a large CX contribution ($\gtrsim50~\%$) for the O~He$\alpha$ $f$ if the CX is assumed to occur between the gas and a hot plasma whose temperature $kT_e$ is $\sim0.2$~keV, $\sim0.5$~keV, or $\sim0.9$~keV reported by previous work~\citep{Konami2011,Lopez2020}.
In Figure~\ref{fig4}(b), we plotted the observed flux of O~Ly$\alpha$ overlaid with the expected flux level assuming that CX is responsible for 50\% of the total flux of \ion{O}{7}~He$\alpha$~$f$.
This comparison requires an upper limit for $kT_{e,\,{\rm CX}}\sim0.35~{\rm keV}$ for the plasma component receiving electrons via CX, regardless of the collision velocity.
In following spectral analysis, we thus assume that CX occurs between the gas and the coldest plasma component with $kT_e\sim0.2~{\rm keV}$.

Based on this assumption, we fitted the RGS first- and second-order spectra.
We employed {\tt bvapec} implemented in {\tt XSPEC} to describe the multiple collisional ionization equilibrium plasma components (hot:\,$kT_{e,\,{\rm hot}}\sim0.9$~keV; medium hot:\,$kT_{e,\,{\rm med}}\sim0.5$~keV; cold:\,$kT_{e,\,{\rm cold}}\sim0.2$~keV).
To model the CX emission, we used {\tt ACX2}\footnote{https://github.com/AtomDB/ACX2} where the CX temperature $kT_{e,\,{\rm CX}}$ is linked to electron temperature $kT_{e,\,{\rm cold}}$ in the cold plasma.
{\tt ACX2} includes any velocity-dependent effects for the calculation of cross section not included in the previous {\tt ACX} model.
The collision velocity $\upsilon_{\rm col}$ in {\tt ACX2} cannot be constrained and does not significantly affect the final results, so we fix $\upsilon_{\rm col}$ at $1000~{\rm km/s}$.
We set the ($n$, $l$) distribution\footnote{https://github.com/AtomDB/ACX2/blob/master/pdf/acx2.pdf} of the exchanged ions to the default value 8, and assumed the case that one ion repeatedly captures electrons until it becomes neutral, which is available in the current CX model.
The electron temperature of the plasma components and the normalization of the plasma and CX components were allowed to vary.
We tied the abundances of C, N, O, Fe, and Ni among both the components and let them vary.
The abundances of metals with line emission not detected were fixed to 1 solar.
For the intrinsic absorption for M82 ($N_{\rm H,\,M82}$) and the galactic absorption ($N_{\rm H,\,MW}$) in the direction toward M82, we used {\tt TBabs} with solar abundances \citep{Wilms2000} and {\tt TBvarabs} with the metal abundances in \cite{Origlia2004}, respectively.
The hydrogen column density $N_{\rm H,\,M82}$ of the former was fixed to $0.4\times10^{21}~{\rm cm^{-2}}$ \citep{Dickey1990}, whereas that of the latter was left as a free parameter.
In addition to these models, we added a power law to account for nonthermal emission from M82 X-1.
The photon index $\Gamma$ was fixed at 0.55 given by \cite{Konami2011} whereas the normalization was left free.
To account for the variation of the flux profile with energy range, we applied different 5~RMFs convolved with the five energy band images of 0.45--0.60~keV, 0.60--0.70~keV, 0.70--0.85~keV, and 0.85--1.25~keV.
We did not use the spectra of the energy band above 1.25~keV as the significant difference of the profiles of the thermal and nonthermal emission creates a large uncertainty in the RMFs.
This modeling gives good fits, as shown in Figure~\ref{fig5}.
The best-fit parameters are summarized in Table~\ref{table3}.

\section{Discussion}

\begin{figure}[ht]
\begin{center}
\vspace{-0mm}
\includegraphics[width=7.5cm]{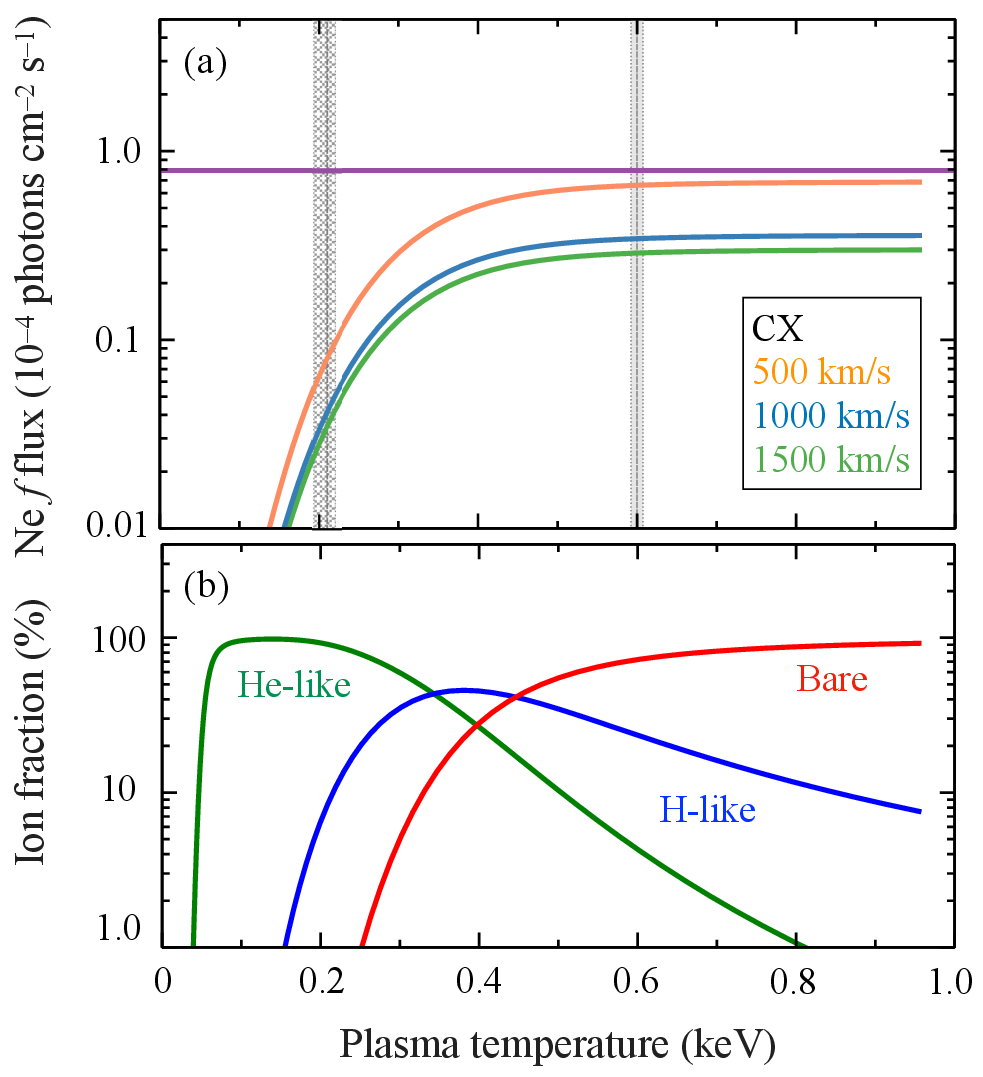} 
\end{center}
\vspace{-5mm}
\caption{
(a) Comparison between the observational \ion{Ne}{7} $f$ flux (purple) and that theoretically expected in the CX process with $\upsilon_{\rm col}=500~{\rm km/s}$ (orange), $1000~{\rm km/s}$ (blue), and $1500~{\rm km/s}$ (green), respectively. 
The gray hatched ($\sim0.2$~keV) and filled ($\sim0.6$~keV) areas correspond to the constrained $kT_{e,\,{\rm CX}}$ range in our results and in \cite{Zhang2014}, respectively. 
(b) Ion fraction of Ne in the He-like (green), H-like (blue), and bare state (red) as a function of the plasma temperature.
}
\label{fig6}
\end{figure}

\begin{figure}[ht]
\begin{center}
\vspace{-0mm}
\includegraphics[width=7.5cm]{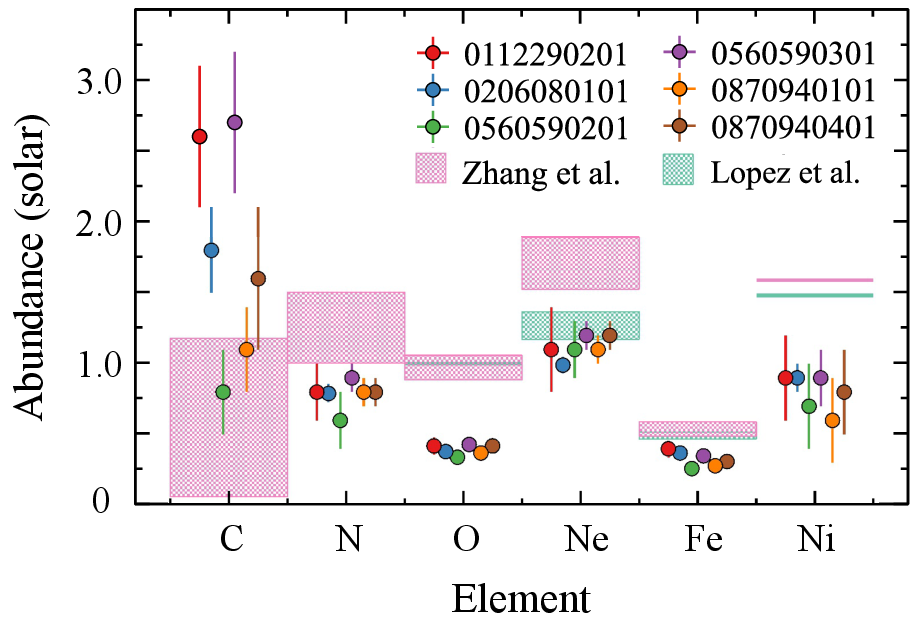} 
\end{center}
\vspace{-5mm}
\caption{
Elemental abundances relative to the solar values of \cite{Wilms2000}.
The magenta and light green hatched areas correspond to previous measurements in \cite{Zhang2014} and Region S1 by \cite{Lopez2020}, that are also normalized by the solar abundances of \cite{Wilms2000} and plotted for comparison.
}
\label{fig7}
\end{figure}

\begin{figure*}[ht]
\begin{center}
\vspace{-0mm}
\includegraphics[width=15cm]{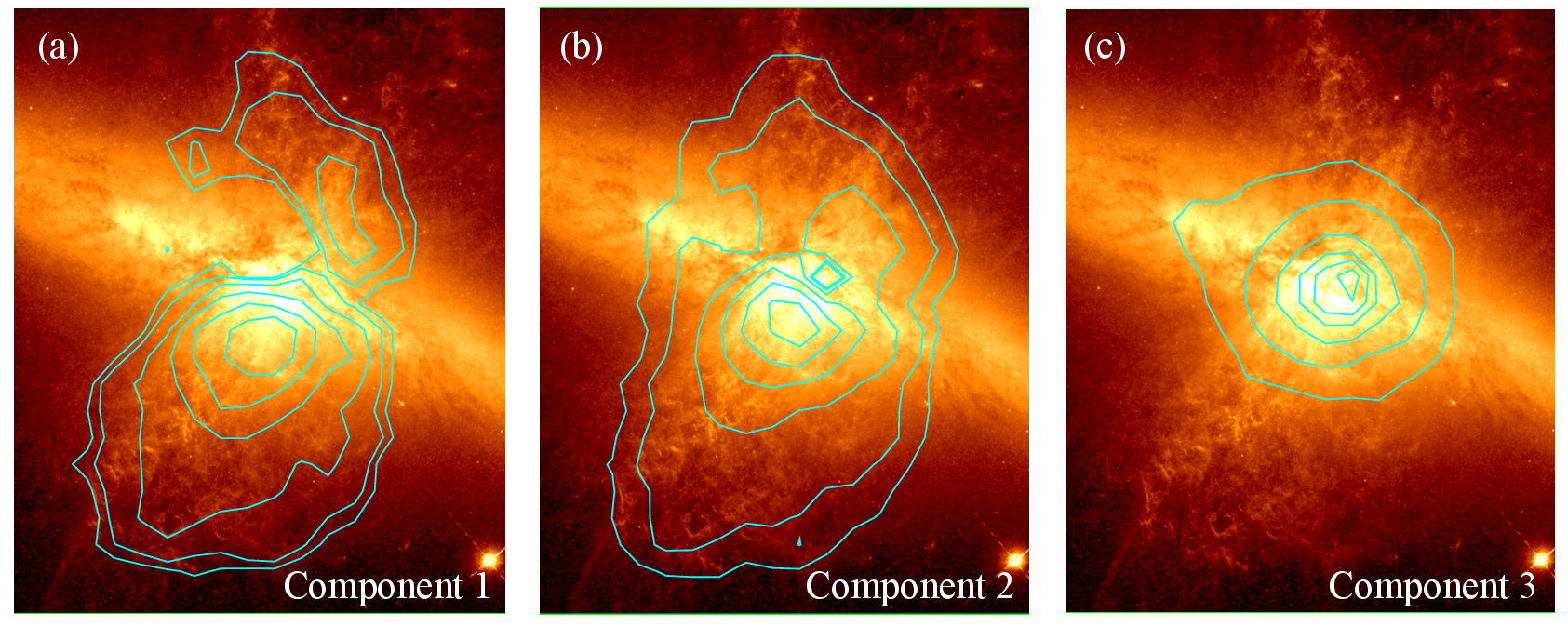} 
\end{center}
\vspace{-5mm}
\caption{
H$\alpha$ band image (F658N) taken with ACS onboard HST \citep{Mutchler2007}.
The cyan contours in the panels (a), (b), and (c) indicate the counts maps of the components~1, 2, and 3 in Figure\ref{fig1}(b), respectively
}
\label{fig8}
\end{figure*}

\subsection{CX contribution}

We have applied the GMCA method to the CCD data of the XMM observations of M82, and for the first time constrained the location of the component with enhanced CX emission in the starburst galaxy. 
Analyzing the RGS spectra from the Peak region in the south outflow, we have detected unusually high \ion{O}{7}~He$\alpha$ $f/r$ ratio as well as clearly resolved emission lines selectively enhanced in the CX process such as \ion{C}{6}~K$\gamma$, \ion{N}{7}~Ly$\alpha$, and \ion{O}{8}~Ly$\alpha$ .
The high ratio and the presence of line emission using parts of the RGS we used have been previously discussed \cite[e.g.,][]{Liu2011,Zhang2014}.
Our results confirm their conclusion and indicate that the CX emission accounts for $\sim50\%$, $\sim30\%$, $\sim60\%$, $\sim40\%$, and $\sim30\%$ of the total flux of the \ion{C}{6}~K$\gamma$, \ion{N}{7}~Ly$\alpha$, \ion{O}{7} $f$, \ion{O}{7} $r$, and \ion{O}{8}~Ly$\alpha$ lines, respectively.

Some previous work \cite[e.g.,][]{Zhang2014,Lopez2020} discussed the CX contribution in the \ion{Ne}{7} and \ion{Mg}{9} triplets in addition to the above line emission.
For example, \cite{Zhang2014} claimed the CX is responsible for more than $30\%$ of the total flux of \ion{Ne}{7}~$f$ although the contribution in our spectral fits is less than $\sim$10\%.
The discrepancy between these and our results arises from the selection of the recipient plasma in the CX modeling.
Figure~\ref{fig6}(a) shows the Ne $f$ flux directly estimated with RGS data via the manner\footnote{
To measure the Ne $f$ and $r$ flux, we delete the two transition in the XSPEC model package and perform the spectral fittings with the best-fit model in Table~\ref{table3} plus additional Gaussians accounting for the two lines. Detailed procedures are described in Sec~5 in \cite{Suzuki2020}} used in \cite{Suzuki2020} and the flux level expected the CX process as a function of the plasma temperature.
In the flux calculation, we have assumed that 60\% of the observed \ion{O}{7}~$f$ line is emitted in the CX process and computed the flux ratio of \ion{Ne}{9}~$f$\,/\,\ion{O}{7}~$f$ with the same package in Sec~\ref{sec:SA} using the abundance ratio $Z_{\rm O}/Z_{\rm Ne}=1.1/0.4$ from Table~\ref{table2}.
The Ne~$f$ flux curve significantly varies with the plasma temperature since the population of H-like and bare ions that can emit the Ne $f$ lines after the electron transfer is sensitive to the plasma temperature (Figure~\ref{fig6}(b)).
If we employ $kT_{e,\,{\rm CX}}\sim0.2~{\rm keV}$ as demonstrated in Sec~\ref{sec:SA}, the CX contribution for the Ne~$f$ is less than $\sim$10\%.
On the other hand, \cite{Zhang2014} assume $kT_{e,\,{\rm CX}}$ to be $\sim0.6~{\rm keV}$, suggesting the CX contribution is overestimated.

This change in the modeled CX emission implies that the abundance measurements in the hot plasmas must be updated as well.
In Figure~\ref{fig7}, we compare the obtained abundance pattern with that given by \cite{Zhang2014} and \cite{Lopez2020}.
It is worth pointing out that overall our abundances tend to be lower except for carbon showing the opposite trend even if we consider their differences between observations.
The cause of the abundance variations could be explained by the difference of the source regions due to the different roll angles and/or the position-to-position fluctuation of the column density, although it is hard to conclude it.
A collection of Core Collapse (CC) events arising from star formation would naturally lead to super-solar abundance ratios of $\alpha$ elements versus Fe \citep[e.g.,][]{Nomoto2006}.
Although the [Ne/Fe]$\,(=\log_{10}(Z_{\rm Ne}/Z_{\rm Fe}))\sim0.51$ ratio remains high and consistent with typical nucleosynthesis models, the [O/Fe]$\sim0.07$ is much smaller.
Given that \cite{Origlia2004} reported high ratio of $\alpha$ elements including O to Fe with the near-infrared observations of stars and cold gas in the disc, the puzzling [O/Fe] behavior should be interpreted as the result of significant O depletion rather than Fe enhancement.
Some O might suffer severe dust depletion since CC supernovae produce the large amount of dust \citep[e.g.,][]{Todini2001}.
However, in the scenario, other elements are also expected to be depleted \citep[e.g.,][]{Savage1996,Boogert2015} so that it is needed to confirm if the observed abundance pattern can be explained by the model including dust evolution.
An alternative possibility is the uncertainty due to the modeling (e.g., are abundances the same among the three plasma and CX components? or are $kT_{ e,\,\rm CX}$ and $kT_{ e,\,\rm cold}$ really the same?).
In either case, follow-up observation with micro-calorimeters onboard future satellites such as {\it XRISM}, will offer better understanding of the chemical property of the starburst galaxy.

\subsection{CX geometry}

The maps of the three components extracted in the GMCA analysis and their comparison with the gas distribution provides an unbiased look at the geometry of the CX component of the galaxy.
Figure~\ref{fig8} shows the location of each component overlaid on the H$\alpha$ image with the Advanced Camera for Surveys (ACS) onboard the Hubble Space Telescope (HST) \citep{Mutchler2007}.
We find Component~1 with enhanced CX emission to be spatially coincident with the distribution of  the H$\alpha$ filaments extending in both north and south sides of the galactic disk.
Numerical \citep[e.g.,][]{Cooper2008}  and hydrodynamical simulations \citep[e.g.,][]{Melioli2013,Schneider2020} show that these vast H$\alpha$ structures are created by swept-up gas that has been dragged by plasma outflows.
Our results support the idea that the CX process occurs at the interface between the plasma and gas components within outflow; similar conclusions were reached by \cite{Lallement2004} and \cite{Wu2020} from numerical simulations.

Another clue to the CX geometry comes from the fraction of the CX emitting volume $V_{\rm CX}$ with respect to the outflow volume $V_{\rm pl}$.
We can estimate $V_{\rm CX}$ to be $3.6\times10^{56}~{\rm cm^{3}}$ (see Appendix~\ref{appendixA}), while $V_{\rm pl} \sim 2.1\times10^{65}~{\rm cm^{3}}$ assuming the morphology to be a cylinder with a radius of 1~kpc ($=1'$) and a height of 2~kpc ($=2'$).
The fraction $\sim10^{-9}$ is significantly smaller than the typical filling factor of the clouds within outflow \citep[$0.001-0.1:$][]{MullerSanchesz2013,Sharp2010}, indicating the CX occurs in an extremely limited region near the interface zone.
This interpretation is indirectly supported by the fact that the temperature $kT_{e,\,{\rm CX}}\sim0.2$~keV of the plasma acting as an electron receiver for the CX process is much lower than the other plasma temperatures ($kT_{e,\,{\rm med}}\sim0.5$~keV and $kT_{e,\,{\rm hot}}\sim0.9$~keV) which account for most of thermal X-rays.
Neutral hydrogen, typically the major electron donor in the CX process, is easily ionized so that it cannot deeply penetrate into the hot plasma \citep[e.g.,][]{Wise1989}.
The low $kT_{e,\,{\rm CX}}$ can be explained if the CX process mainly occurs in the layer with plasma temperature rapidly varying due to thermal conduction by the gas.

\section{Conclusions}

We have performed an analysis of X-ray emission of M82, a prototype starburst galaxy, obtained with {\it XMM-Newton} observations in order to investigate the CX flux contribution and its geometry.
We have applied a blind source separation method to the analysis of the CCD data and identified the spatial distribution of the component wth the enhanced O-K lines expected from the CX process in the starburst galaxy for the first time.
Based on the image analysis, we have analyzed the RGS data extracted from the compact peak ($\sim2'$) in the south outflow and detected a high forbidden-to-resonance ratio in the \ion{O}{7} He$\alpha$ triplet as well as several emission lines enhanced in the CX process such as \ion{C}{5} K$\gamma$, and \ion{N}{7} and \ion{O}{8} Ly$\alpha$.
The RGS spectra of all observations are well fitted with the model that consists of three different plasma temperatures ($\sim0.2$~keV, $\sim0.5$~keV, and $\sim0.9$~keV) and CX components with an additional nonthermal component.
The CX emission accounts for $\sim50\%$, $\sim30\%$, $\sim60\%$, $\sim40\%$, and $\sim30\%$ of the total flux of \ion{C}{6}, \ion{N}{7}, \ion{O}{7}~$f$, \ion{O}{7}~$r$, and \ion{O}{8}~Ly$\alpha$ lines, respectively, although the CX contributions to the emission lines of Ne and Mg are less than 
$\sim$10\%.
Including this CX emission component primarily affects the measured abundance measurement of these light elements in the thermal plasma components, tending towards lower values relative to earlier calculations.
The temperature of the plasma as electron receiver in the CX process is significantly lower than the plasma components that emit most of X-rays.
From the low temperature and an estimation of the CX emitting volume, the CX primarily occurs in a thin region near the interface of the plasma and gas whose temperature rapidly decreases due to thermal conduction.

\acknowledgments

We deeply appreciate all the {\it XMM-Newton} and HST members.
We are also grateful to Drs. John C. Raymond and Q. Daniel Wang for helpful discussion and Dr. Caroline Huang for analysis of HST data.
H.O. is supported by Japan Society for the Pro-motion of Science (JSPS) Research Fellowship for Research Abroad.

\appendix
\section{Estimation of $V_{\rm CX}$} \label{appendixA}
The CX volume $V_{\rm CX}$ can be estimated from the emission measure of the CX component VEM$_{\rm CX}$, defined as $\int n^{d}_{\rm (H+He)}\,n^{r}_{\rm H}\,dV_{\rm CX}$, where $n^{d}_{\rm (H+He)}$ and $n^{r}_{\rm H}$ are the total density of neutral hydrogen and helium of the donor gas and the hydrogen density of the receiver plasma, respectively.
The average VEM$_{\rm CX}$ is estimated to be $0.9\times10^{57}~{\rm cm^{-3}}$ by applying Equation \ref{eq1} to values in Table~\ref{table2}.
In order to calculate $V_{\rm CX}$, we need to obtain $n^{d}_{\rm (H+He)}$ and $n^{r}_{\rm H}$.
According to \cite{Shopbell1998}, $n^{d}_{\rm (H+He)}$ is $\sim14~{\rm cm^{-3}}$ under the assumption of solar abundances of H and He.
The $n^{r}_{\rm H}$ is given by the emission measure of the plasma components ${\rm EM}_{\rm pl}$, defined as $\int n^{r}_{\rm H}\,n^{r}_{e}\,dV_{\rm pl}~{\rm cm^{-3}}$, where $n^{r}_{e}$ is the electron density of the receiver plasma.
Assuming the average ${\rm VEM}_{\rm pl} = {\rm VEM}_{\rm cold} + {\rm VEM}_{\rm med} +{\rm VEM}_{\rm hot} = 8.1\times10^{61}~{\rm cm^{-3}}$, we can estimate $n^{r}_{\rm H}$ to be $\sim0.018~{\rm cm^{3}}$.
Here, we have used the relation of $1.2\,n^{r}_{\rm H}\sim n^{r}_{e}$.
Finally, we obtain V$_{\rm CX}$ to be $3.6\times10^{57}{\rm cm^{3}}$.

\end{document}